\documentclass{moriond}

\usepackage{csquotes}
\usepackage[capitalize]{cleveref}
\usepackage{amssymb}
\usepackage{siunitx}

\newcommand{\msbar}{$\overline{\rm { MS}}$}
\newcommand{\mmsbar}{{\overline{\rm {MS}}}}

\newcommand{\Photo}{}

\begin{document}
\vspace*{4cm}
\title{On the positivity of \msbar{} distributions}

\author{ Felix Hekhorn }

\address{University of Jyvaskyla, Department of Physics, P.O. Box 35, FI-40014 University of Jyvaskyla, Finland\\
Helsinki Institute of Physics, P.O. Box 64, FI-00014 University of Helsinki, Finland}

\maketitle\abstracts{
       We discuss the positivity of parton distribution functions using the common \msbar{} factorization scheme.
       We find that in the perturbative regime \msbar{} PDFs inherit the strict positivity of physical PDFs.
       We explicitly discuss the scheme transformation by using suitable physical observables and find that
       \msbar{} PDFs are positive above the scale $Q^2 \gtrsim \SI{5}{\GeV^2}$.
       Finally, we comment on the direct counterpart of longitudinally polarized PDFs, finding agreement with the
       unpolarized counterpart.
}

\section{Unpolarized PDFs}

The positivity\footnote{Here and in the following by \enquote{positive} we mean more precisely \enquote{non-negative}.} of parton distribution
functions (PDF) was a long-standing issue in the community of PDF fitting groups. If and how positivity should or should not be enforced
onto the parametrization of PDF is of immediate consequence onto resulting PDF and different PDF fitting groups have made different choices.
While the positivity of physical observables follows from unitarity arguments on the theory side, the positivity of PDFs can not be
directly deduced from this constrain. However, the inverse argument does hold: a PDF which results in negative physical observables
signals a problem in the PDF definition.

We addressed the issue of positivity directly in our first paper\cite{Candido:2020yat} and provided arguments that within the framework
of leading-twist expansion, in which PDFs are defined, in a perturbative regime \msbar{} PDFs are positive.
We argued that the positivity of physical PDFs is inherited by the more commonly used \msbar{} PDFs.
While in our first paper\cite{Candido:2020yat} we constructed a dedicated factorization scheme, the so-called
\enquote{positivity scheme}, to prove the positivity, we realized in our second paper\cite{Candido:2023ujx} that
we can argue directly starting from a so-called physical scheme\cite{Diemoz:1987xu,Catani:1995ze,Catani:1996sc,Altarelli:1998gn,Lappi:2023lmi},
which we briefly outline here below.

In a physical PDF scheme, physical observables are \textit{identified} with PDFs and the
prototype of such a scheme is the DIS scheme\cite{Diemoz:1987xu}, in which the $F_2$ structure
function in deep inelastic scattering (DIS) is given by the quark PDFs:
\begin{equation}
       F_2(x,Q^2) = \sum_q e_q^2 f_q^{\rm DIS}(x,Q^2) + O(\Lambda^2 /Q^2).\label{eq:DIS}
\end{equation}
Note that the DIS scheme is still one possible choice for a PDF scheme, i.e.\ it is only valid in the
leading-twist expansion and we explicitly indicated the suppressed terms which are beyond the collinear
factorization framework.

\cref{eq:DIS} can be generalized to all parton flavors and we can write for a set of physical observables $\sigma$
\begin{equation}
       \sigma(x,Q^2) = \sigma_0 f^{\rm PHYS}(x,Q^2) + O(\Lambda^2 /Q^2) \label{eq:phys}
\end{equation}
given a suitable normalization $\sigma_0$.
Hence, PDFs in this physical factorization scheme are positive by definition.
The direct counterpart to \cref{eq:phys} in the \msbar{} factorization scheme, which is employed by most modern PDF fitting groups,
is given by
\begin{equation}
       \sigma(x,Q^2) = \sigma_0 C^{\mmsbar}(\alpha_s(Q^2), Q^2/\mu_F^2) \otimes f^{\mmsbar}(\mu_F^2) + O(\Lambda^2 /Q^2) \label{eq:msbar}
\end{equation}
with the universal coefficient function $C^{\mmsbar}$ which is computed perturbatively and hence depends explicitly on the
strong coupling $\alpha_s$ and the factorization scale $\mu_F$.

Comparing \cref{eq:DIS,eq:msbar} with each other we can deduce the perturbative scheme transformation between the PDF in the physical
scheme and the PDF in the \msbar{} scheme and we find
\begin{equation}
       f^{\mmsbar}(x,Q^2) = \left[C^{\mmsbar}(\alpha_s(Q^2))\right]^{-1} \otimes f^{\rm PHYS}(Q^2) \label{eq:trafo}
\end{equation}
where the inverse is to be understood as acting in a multiplicative distribution sense and the equation contains a non-trivial flavor
decomposition. Thus, we conclude that \textit{if} the convolution with the inverse coefficient function preserves positivity
the \msbar{} PDF $f^{\mmsbar}$ will inherit the positivity of the physical PDF $f^{\rm PHYS}$.

The perturbative expansion of the \msbar{} coefficient function, $C^{\mmsbar}$, is given by
\begin{equation}
       C^{\mmsbar}_{ij}(x) = \delta_{ij}\delta(1-x) +\frac{\alpha_s}{2\pi}\left[ 
              \delta_{ij} \delta(1-x) \Delta_i^{(1)}+C_{F,ij}^{(1),\,\mmsbar}(x)+C_{D,ij}^{(1),\,\mmsbar}(x)
       \right] + \mathcal{O}(\alpha_s^2) \label{eq:cexp}
\end{equation}
where we impose (without loss of generality) that the the leading-order (LO) contribution is a delta distribution and we decompose
the next-to-leading order (NLO) contribution into three terms: a purely local contribution $\Delta^{(1)}$, a purely
plus-distributional contribution $C_D^{(1),\,\mmsbar}$ and the remaining finite contribution $C_F^{(1),\,\mmsbar}$.
Inserting \cref{eq:cexp} into \cref{eq:trafo} it can be shown that the positivity condition can be satisfied perturbatively
which in turn requires that the convolution of $C_F^{(1),\,\mmsbar}$ with the PDF is smaller in magnitude than the PDF itself.
Straightforward manipulations lead to the conclusion that this is true provided that
\begin{equation}\label{eq:poscondfin}
\frac{\alpha_s}{2\pi} \left| \sum_j  {\cal C}_{ij}(x) \right| \le 1,
\end{equation}
where we have defined cumulants
\begin{equation}\label{eq:cumulant}
{\cal C}_{ij}(x)  = \int_x^1\frac{dy}{y} C^{(1),\mmsbar}_{F,ij}(y) \, .
\end{equation}

\begin{figure}
\begin{minipage}{0.45\linewidth}
\centerline{\includegraphics[width=0.7\linewidth]{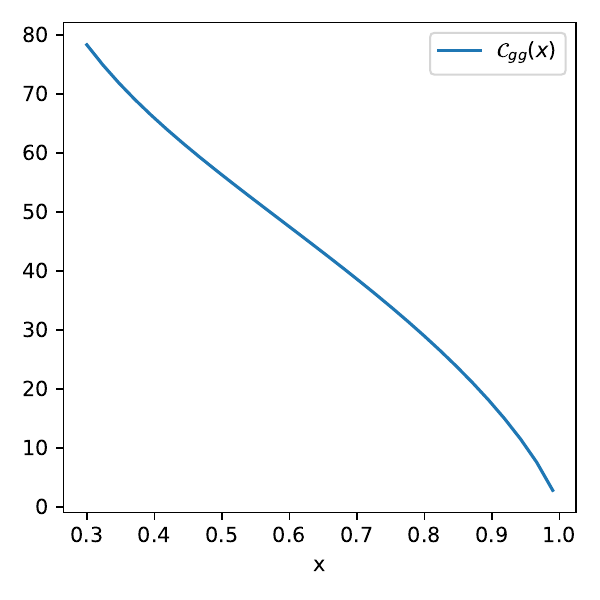}}
\end{minipage}
\begin{minipage}{0.45\linewidth}
\centerline{\includegraphics[width=0.7\linewidth]{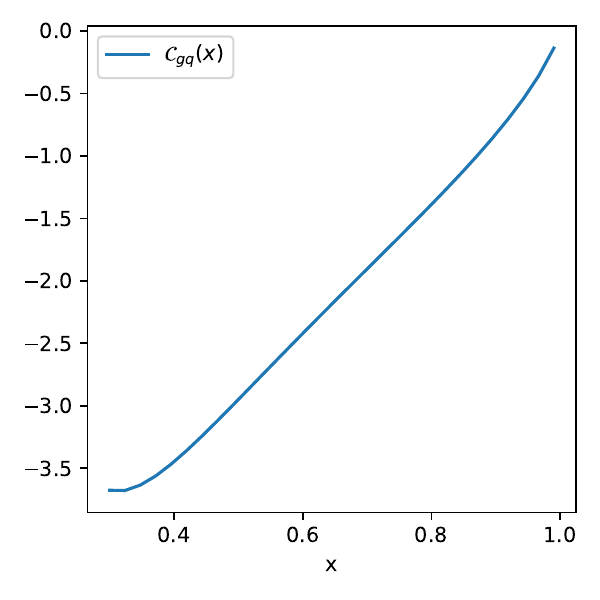}}
\end{minipage}

\begin{minipage}{0.45\linewidth}
\centerline{\includegraphics[width=0.7\linewidth]{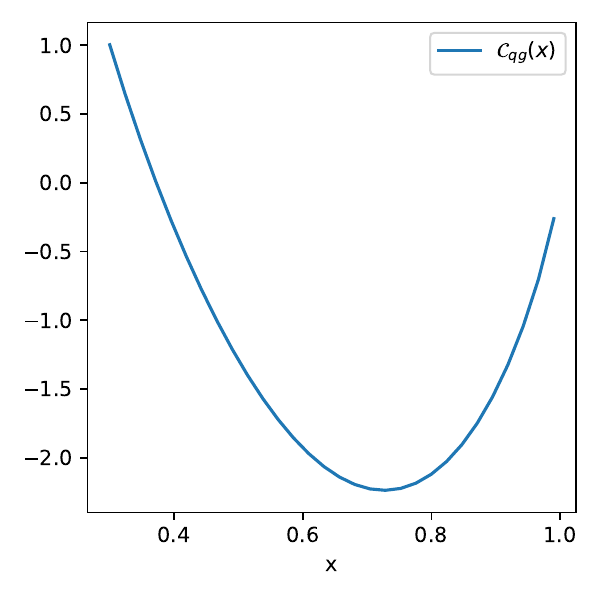}}
\end{minipage}
\begin{minipage}{0.45\linewidth}
\centerline{\includegraphics[width=0.7\linewidth]{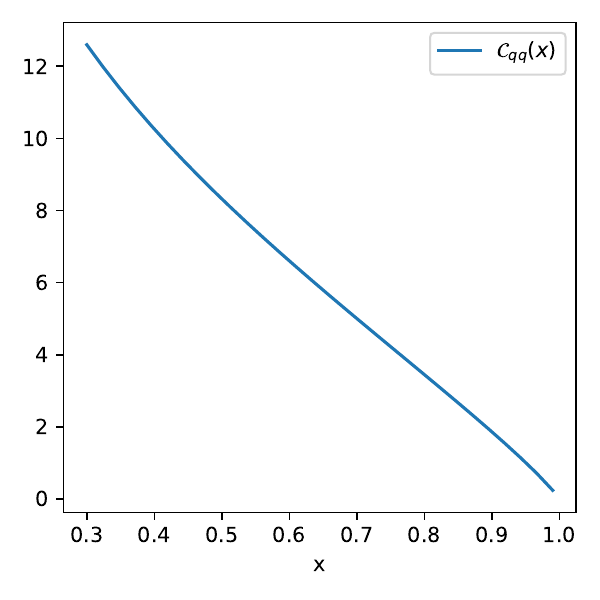}}
\end{minipage}
\caption{Unpolarized cumulants $\mathcal C_{ij}(x)$}
\label{fig:unpol}
\end{figure}

Choosing DIS and Higgs production in gluon fusion as definition for the physical observables, as suggested by Ref.\cite{Altarelli:1998gn},
we obtain \cref{fig:unpol} as cumulants for the respective flavor combinations.
By imposing \cref{eq:poscondfin} we find as a lower bound $Q^2 \gtrsim \SI{5}{\GeV^2}$ above which positivity should be imposed.
Note that in our recent paper\cite{Candido:2023ujx} we confirm via an explicit calculation the issue raised in Ref.\cite{Collins:2021vke}
that at sufficiently low scales the \msbar{} PDF will turn negative.

\section{Polarized PDFs}

\begin{figure}
\begin{minipage}{0.45\linewidth}
\centerline{\includegraphics[width=0.7\linewidth]{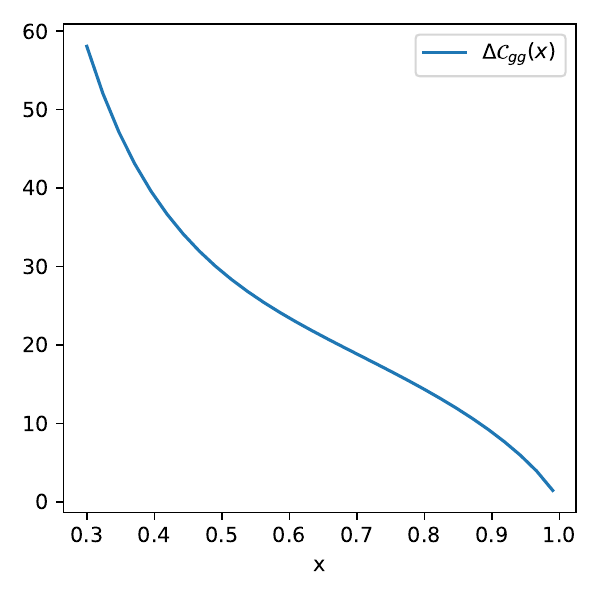}}
\end{minipage}
\begin{minipage}{0.45\linewidth}
\centerline{\includegraphics[width=0.7\linewidth]{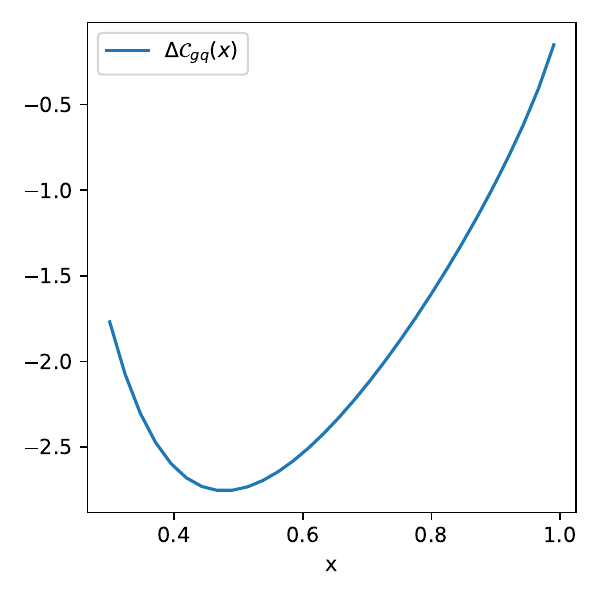}}
\end{minipage}

\begin{minipage}{0.45\linewidth}
\centerline{\includegraphics[width=0.7\linewidth]{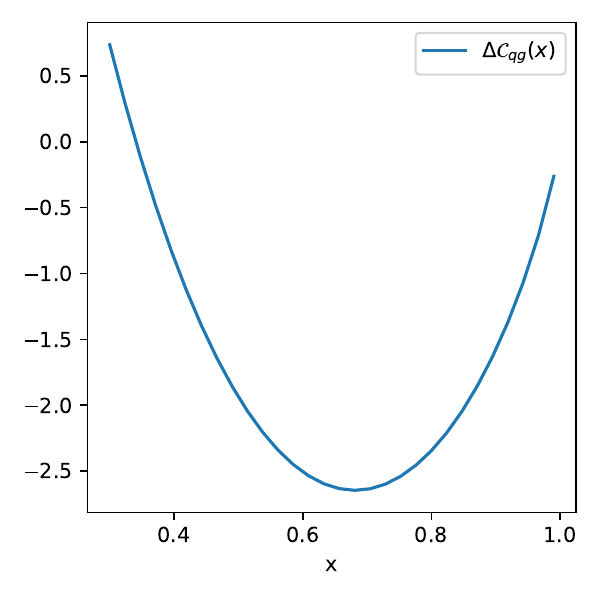}}
\end{minipage}
\begin{minipage}{0.45\linewidth}
\centerline{\includegraphics[width=0.7\linewidth]{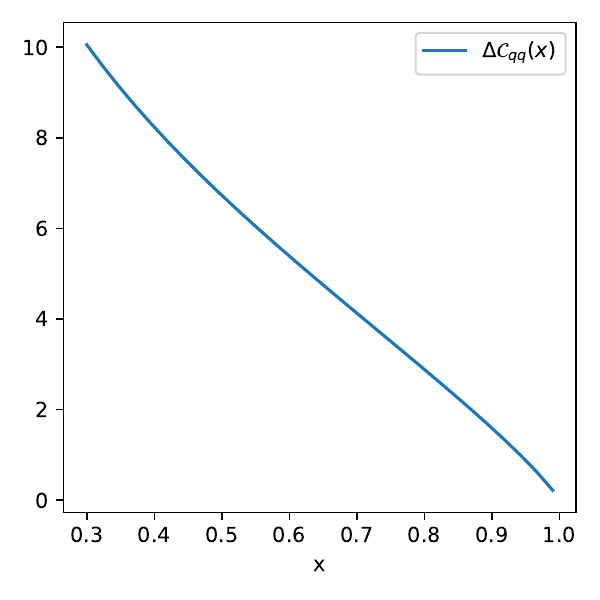}}
\end{minipage}
\caption{Longitudinally polarized cumulants $\Delta \mathcal C_{ij}(x)$}
\label{fig:pol}
\end{figure}

The above discussion holds independently of the spin-alignment of the scattering parton and its parent hadron
and we now comment on the case of longitudinally polarized PDFs. Helicity-dependent PDFs $\Delta f$ and their unpolarized
counterparts $f$ are defined as the difference and sum of individual spin contributions respectively via
\begin{equation}
       f = f^{\uparrow\uparrow} + f^{\uparrow\downarrow}\quad \text{and} \qquad \Delta f = f^{\uparrow\uparrow} - f^{\uparrow\downarrow}
\end{equation}
where the arrows indicate the alignment of the hadron and parton respectively.
As polarized PDFs and their corresponding physical observables are defined by a difference their positivity condition
refers to their \textit{absolute} value instead.
Starting from the physical observable constrain
\begin{equation}
       0 \leq |\Delta \sigma| \leq \sigma
\end{equation}
one can deduce\cite{Altarelli:1998gn} the corresponding condition for the PDFs:
\begin{equation}
       0 \leq |\Delta f| \leq f \label{eq:polpos}
\end{equation}
In order for the unpolarized PDF $f$ to be positive we have to impose the constrain that is deduced in this
context of $Q^2 \gtrsim \SI{5}{\GeV^2}$.
However, it could be that the perturbativity condition, \cref{eq:poscondfin}, becomes more restrictive in the polarized case.
In order to check this, we compute the polarized cumulants, obtained by replacing the unpolarized coefficient functions in \cref{eq:cumulant}
with their polarized counterpart. The results, displayed in \cref{fig:pol}, shows that polarized and unpolarized cumulants have the same behavior,
so the positivity bound $Q^2 \gtrsim \SI{5}{\GeV^2}$ is the same in the unpolarized and polarized case.

This conclusion is in agreement with the physical observation that near threshold, i.e.\ $x\to 1$,
polarized and unpolarized coefficient functions have the same behavior, which is enforced by
the closing of the available phase space.
       
\section*{Acknowledgments}

F.~H. is supported by the Academy of Finland project 358090 and is funded as a
part of the Center of Excellence in Quark Matter of the Academy of Finland, project 346326.

\end{document}